\documentclass[sigconf]{acmart}
\setcopyright{none}
\renewcommand\footnotetextcopyrightpermission[1]{}
\copyrightyear{2026}
\acmYear{2026}
\acmConference[Preprint]{Preprint}{2026}{arXiv}
\acmBooktitle{Preprint, 2026, arXiv}
\acmDOI{}
\acmISBN{}
\usepackage{algorithm}
\usepackage{algpseudocode}
\usepackage{multirow}
\usepackage{booktabs}
\usepackage{tabularx}

\newcommand\blfootnote[1]{%
  \begingroup
  \renewcommand\thefootnote{}%
  \footnotetext{#1}%
  \addtocounter{footnote}{-1}%
  \endgroup
}

\newcommand{\equalcontrib}{\textsuperscript{*}}
\newcommand{\corrauthor}{\textsuperscript{\ensuremath{\dagger}}}

\AtBeginDocument{%
  }

\begin{document}

\title{Beyond Isolated Utterances: Cue-Guided Interaction for Context-Dependent Conversational Multimodal Understanding}

\author{Zhaoyan Pan\equalcontrib}
\email{zhaoyanpan@zju.edu.cn}
\affiliation{%
  \institution{Zhejiang University}
  \city{Hangzhou}
  \country{China}
}

\author{Hengyang Zhou\equalcontrib}
\email{hengyangzhou@smail.nju.edu.cn}
\affiliation{%
  \institution{Nanjing University}
  \city{Nanjing}
  \country{China}
}

\author{Xiangdong Li}
\email{xiangdong.li@zju.edu.cn}
\affiliation{%
  \institution{Zhejiang University}
  \city{Hangzhou}
  \country{China}
}

\author{Yuning Wang}
\email{yuning.wang@zju.edu.cn}
\affiliation{%
  \institution{Zhejiang University}
  \city{Hangzhou}
  \country{China}
}

\author{Ye Lou}
\email{ye_lou@zju.edu.cn}
\affiliation{%
  \institution{Zhejiang University}
  \city{Hangzhou}
  \country{China}
}

\author{Jiatong Pan}
\email{jiatongpan@zju.edu.cn}
\affiliation{%
  \institution{Zhejiang University}
  \city{Hangzhou}
  \country{China}
}

\author{Ji Zhou}
\email{jizhou@zju.edu.cn}
\affiliation{%
  \institution{Zhejiang University}
  \city{Hangzhou}
  \country{China}
}

\author{Wei Zhang\corrauthor}
\email{cstzhangwei@zju.edu.cn}
\affiliation{%
  \institution{Zhejiang University}
  \city{Hangzhou}
  \country{China}
}

\renewcommand{\shortauthors}{Pan, Zhou et al.}

\begin{abstract}
Conversational multimodal understanding aims to infer the meaning or label of the current utterance from its preceding dialogue context together with textual, acoustic, and visual signals. Existing methods mainly strengthen contextual modeling through enhanced encoding, fusion, or propagation, but rarely abstract the context-utterance dependency into an explicit cue and incorporate it into later multimodal reasoning. To address this issue, we propose CUCI-Net for conversational multimodal understanding. CUCI-Net fully preserves the structural distinction between context and utterance during encoding, effectively abstracts their dependency into an interpretation cue by combining local modality evidence with global contextual evidence, and seamlessly integrates the resulting cue into the final multimodal interaction stage for context-conditioned prediction. Extensive experiments on mainstream benchmark datasets fully demonstrate the effectiveness of the proposed method.
\end{abstract}

\begin{CCSXML}
<ccs2012>
   <concept>
       <concept_id>10010147.10010178.10010179.10010181</concept_id>
       <concept_desc>Computing methodologies~Discourse, dialogue and pragmatics</concept_desc>
       <concept_significance>500</concept_significance>
   </concept>
</ccs2012>
\end{CCSXML}

\ccsdesc[500]{Computing methodologies~Discourse, dialogue and pragmatics}

\keywords{multimodal learning, multimodal sentiment analysis, conversational understanding, multimodal fusion}

\maketitle

\blfootnote{\equalcontrib{} Equal contribution. \quad \corrauthor{} Corresponding author.}

\section{Introduction}

Conversational multimodal understanding aims to predict the meaning or label of the current utterance from its preceding dialogue context together with textual, acoustic, and visual signals \cite{poria-etal-2019-meld,castro-etal-2019-towards,hasan-etal-2019-ur-funny,hu-etal-2022-unimse}. Unlike sentence-level multimodal prediction, the current utterance in a conversation is often ambiguous in isolation: its sentiment, humorous effect, or non-literal intent may only become clear after considering how the preceding discourse frames or shifts its meaning \cite{hazarika-etal-2018-conversational,ghosal-etal-2019-dialoguegcn,hu-etal-2021-dialoguecrn,zhang-li-2023-cross}. Meanwhile, facial behaviors and vocal expressions further modulate whether the utterance should be interpreted as sincere, ironic, exaggerated, or emotionally shifted \cite{castro-etal-2019-towards,hasan-etal-2019-ur-funny,zheng-etal-2023-facial}. Therefore, the challenge lies not only in collecting multimodal signals, but in preserving the context-utterance dependency and using this dependency to guide the final interpretation. This is especially central in conversational benchmarks, where the label of the current utterance is often determined more by context-utterance dependency than by the utterance alone \cite{poria-etal-2019-meld,castro-etal-2019-towards,hasan-etal-2019-ur-funny}.

\begin{figure}[t]
  \centering
  \includegraphics[width=0.9\linewidth]{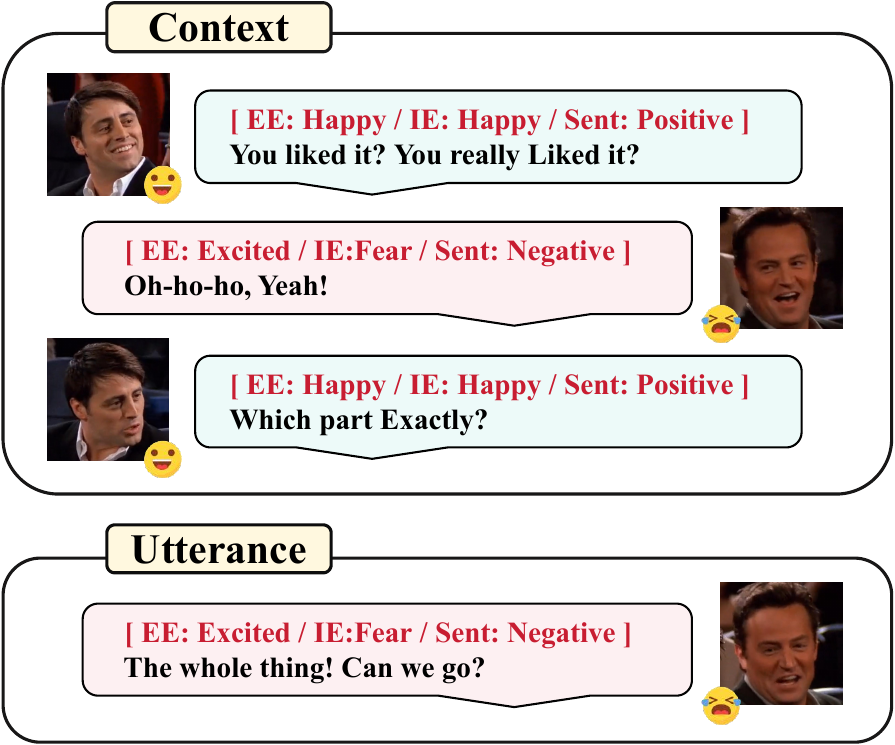}
  \caption{An example where the current utterance can only be correctly understood with the help of its preceding conversational context. EE, IE, and Sent denote Explicit Emotion, Implicit Emotion, and sentiment, respectively.}
  \label{fig:dual_expert}
  \vspace{-0.2cm}
\end{figure}

Prior studies have shown the importance of contextual modeling for conversational understanding \cite{hazarika-etal-2018-conversational,ghosal-etal-2019-dialoguegcn,hu-etal-2021-dialoguecrn,zhang-li-2023-cross,shi-huang-2023-multiemo,zheng-etal-2023-facial}. Recent multimodal methods improve contextual modeling through enhanced fusion, graph-based interaction, or auxiliary supervision \cite{yun-etal-2024-telme,hossain-etal-2024-m3tcm,tu-etal-2024-multiple,ai2025graphspectrum,zhou2025robustrealiblemultimodalmisinformation,zhou2026diverdynamiciterativevisual}. However, most of them still strengthen context modeling mainly through latent encoding, propagation, or fusion. Even when contextual structure is modeled more carefully, what is preserved is usually contextual content itself, rather than an explicit characterization of how the context conditions the interpretation of the current utterance \cite{yun-etal-2024-telme,hossain-etal-2024-m3tcm,xue-etal-2024-breakthrough,ai2025graphspectrum}. As a result, the fine-grained context-utterance dependency is rarely abstracted into an explicit interpretation cue, and even less often explicitly incorporated into the subsequent complex multimodal reasoning process to reasonably regulate how later interaction unfolds.

Effective conversational multimodal understanding therefore requires more than stronger context-aware fusion. It requires strictly preserving the context-utterance structure during representation learning, abstracting the resulting dependency into a clear explicit interpretation cue, and incorporating this cue into later multimodal reasoning so that it can shape the final interpretation process rather than remain a passive byproduct of encoding. Such a targeted design enables the model to capture latent semantic constraints and deliver more coherent dialogue comprehension results.

Based on this idea, we propose CUCI-Net, a \textbf{C}ontext-\textbf{U}tterance and \textbf{C}ue-guided \textbf{I}nteraction \textbf{Net}work for conversational multimodal understanding. CUCI-Net follows a three-stage interpretation-cue-guided reasoning framework. It first preserves the structural distinction between the contextual segment and the current utterance during representation learning and derives a text-anchored relation representation for the acoustic and visual modalities. It then constructs the final interpretation cue by combining local modality evidence with global contextual evidence, so that the context-utterance dependency becomes available in a compact form for later reasoning. Finally, it effectively injects this interpretation cue into the subsequent multimodal interaction process, so that the cue not only represents how context conditions the current utterance, but also steers how later multimodal reasoning and interaction evolve toward the accurate final prediction.

Our contributions are summarized as follows:
\begin{itemize}
    \item We propose CUCI-Net, a three-stage cue-guided framework for conversational multimodal understanding that explicitly preserves the context-utterance structure, abstracts their dependency into an interpretation cue, and incorporates this cue into the final multimodal reasoning stage for context-conditioned prediction.
    
    \item We develop a structure-aware modeling scheme that preserves context-utterance distinction during encoding and uses a text-anchored relation representation to guide acoustic and visual representation learning. We further design a cue-guided interaction mechanism that constructs an interpretation cue from global contextual and local modality evidence and injects it into guided multimodal interaction layers for context-conditioned reasoning.
    
    \item Extensive experiments on the MUStARD and MUStARD++ datasets show that CUCI-Net outperforms strong baselines, confirming the effectiveness of the proposed design.

\end{itemize}

\section{Related Work}

\subsection{Multimodal Dialogue Understanding}

Multimodal dialogue understanding has been widely studied in emotion-centric settings, where the goal is to identify the emotion of each utterance from dialogue context together with multimodal behavioral signals. Early work has established the importance of context and speaker interaction through representative paradigms such as conversational memory modeling in ICON~\cite{hazarika-etal-2018-icon}, speaker-aware recurrent reasoning in DialogueRNN~\cite{majumder-etal-2019-dialoguernn}, commonsense-enhanced contextual modeling in COSMIC~\cite{ghosal-etal-2020-cosmic}, and multimodal graph fusion in MMGCN~\cite{hu-etal-2021-mmgcn}. Subsequent studies further strengthen dialogue understanding with topic- or knowledge-aware contextual encoding and graph-based multimodal reasoning~\cite{zhong-etal-2019-ket,zhu-etal-2021-topic,joshi-etal-2022-cogmen,li-etal-2023-joyful,xu2025towards,wei2024g,wei2024towards}. These works consistently show that multimodal dialogue understanding greatly benefits from modeling conversational history, speaker dependency, and cross-modal complementarity rather than naively interpreting each utterance in isolation.

Beyond emotion recognition, multimodal dialogue understanding has also been explored in sarcasm and humor-related tasks, where meaning often arises from contextual incongruity, pragmatic reversal, or subtle audiovisual cues. Representative studies incorporate sentiment and emotion signals for humor understanding~\cite{chauhan-etal-2022-humor}, model humor intensity from multimodal evidence~\cite{alnajjar-etal-2022-humor-intensity}, exploit additional cues such as gaze~\cite{tiwari-etal-2023-gaze}, and introduce prompt-based adaptation, context-aware fusion, or ambiguity-aware incongruity modeling for sarcasm understanding~\cite{jana-etal-2024-camp,xue-etal-2024-breakthrough,li-etal-2025-ambiguity,yuan2025enhancing,zhou2025ldgnet,wei2025deepmsd}. Related efforts also extend multimodal understanding to comic-context scenarios and multilingual humor corpora~\cite{guo-etal-2024-much,baharlouei-etal-2024-comic}. Together, these works firmly reinforce the broader view that dialogue-level multimodal understanding is fundamentally context-sensitive, and that the interpretation of the current utterance often depends on how surrounding discourse deeply reshapes its surface meaning.

\subsection{Context-Aware Multimodal Reasoning}

Another related line of research concerns how multimodal reasoning is organized under structural or contextual constraints. Representative methods explore graph-based multimodal interaction in MMGCN~\cite{hu-etal-2021-mmgcn}, topic- and knowledge-aware contextual modeling~\cite{zhu-etal-2021-topic}, joint fusion with graph contrastive learning in Joyful~\cite{li-etal-2023-joyful}, dual-graph dependency modeling in DualGATs~\cite{zhang-etal-2023-dualgats}, and knowledge-enhanced interaction or explicit reasoning-oriented architectures~\cite{tu-etal-2024-multiple,hong-etal-2024-detectivenn}. These studies strongly suggest that multimodal prediction should not rely on simple flat concatenation alone, but should instead be shaped by dialogue structure, modality interaction, and task-relevant guidance.

Nevertheless, in most existing approaches, such guidance is still realized mainly through latent fusion, graph propagation, contrastive optimization, or encoder-side modulation~\cite{tu-etal-2023-contrastive,tu-etal-2024-multiple}. Even when context-sensitive interaction is strengthened, the context-utterance dependency is rarely abstracted into an explicit interpretation cue for subsequent reasoning. Recent multimodal dialogue models continue to improve dependency modeling, evidence aggregation, and dynamic interaction, such as through dual contrastive learning, dynamic graph evolution, and evidence-cause attention~\cite{xie-etal-2025-dclf,shou-etal-2025-dgode,zhang-etal-2025-ecerc}, yet the final decision is still often made from implicitly fused representations. Our method instead explicitly preserves fine-grained context-utterance structure, derives a text-anchored relation representation, and then systematically summarizes the context-utterance dependency into a reliable interpretation cue that guides later multimodal interaction.

\begin{figure*}[t]
  \centering
  \includegraphics[width=1\textwidth]{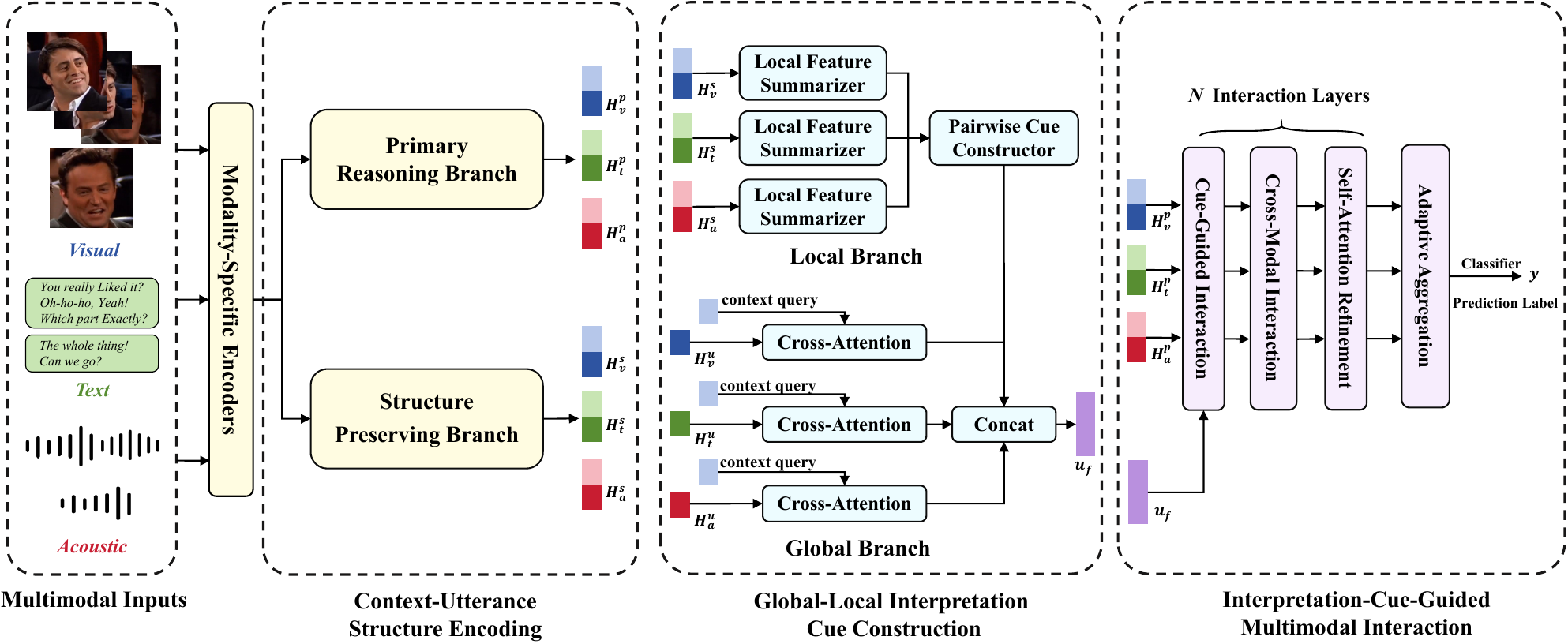}  % 关键修改点
  \caption{CUCI-Net consists of three stages: \emph{Context-Utterance Structure Encoding}, \emph{Global-Local Interpretation Cue Construction}, and \emph{Interpretation-Cue-Guided Multimodal Interaction}.
The first stage learns the primary modality representations $\{H_m^{p}\}_{m\in\{t,a,v\}}$ and the structure-preserving representations $\{H_m^{s}\}_{m\in\{t,a,v\}}$.
The second stage constructs the interpretation cue $u_f$ by combining local pairwise cues with global evidence read out from contextual text.
The third stage injects $u_f$ into stacked multimodal interaction layers for final adaptive prediction.
Blue, green, and red denote the visual, textual, and acoustic modalities, respectively, while light and dark shades denote the context and utterance parts.
}  % 保持原始caption
  \label{fig:architecture}
\end{figure*}

\section{Methodology}

\subsection{Task Formulation}

Let $\mathcal{M}=\{t,a,v\}$ denote the textual, acoustic, and visual modalities.
Each conversational sample is represented as
$\mathcal{X}=\{(C^m,U^m)\mid m\in\mathcal{M}\}$,
where $C^m\in\mathbb{R}^{T_c^m\times d_m}$ and
$U^m\in\mathbb{R}^{T_u^m\times d_m}$ denote the contextual segment and the current utterance of modality $m$, respectively.
Here, $T_c^m$, $T_u^m$, and $d_m$ denote the context length, utterance length, and feature dimension of modality $m$.
For convenience, let $T_m=T_c^m+T_u^m$ denote the length of the joint context-utterance sequence of modality $m$.
The goal is to predict the label $y$ of the current utterance from the multimodal conversational input $\mathcal{X}$.

\subsection{Overview of CUCI-Net}

\textbf{Cue-guided rationale.}
Conversational multimodal understanding requires more than jointly encoding the contextual segment and the current utterance. The contextual segment determines how the current utterance should be interpreted. CUCI-Net therefore follows an interpretation-cue-guided reasoning process: it first preserves context-utterance structure, then abstracts an interpretation cue from this preserved structure, and finally uses the interpretation cue to guide multimodal reasoning over the current utterance.
\par\noindent
\textbf{Three-stage architecture.}
CUCI-Net (Context-Utterance and Cue-guided Interaction Network) comprises three stages: Context-Utter-ance Structure Encoding, Global-Local Interpretation Cue Construction, and Interpretation-Cue-Guided Multimodal Interaction. Given the contextual segment and the current utterance from the textual, acoustic, and visual modalities, the first stage instantiates two architecture-identical but parameter-independent branches. The two branches share the same textual ALBERT encoder, while using separate acoustic and visual Transformers, and serve different downstream roles. One branch provides the primary modality representations for the final reasoning stage, while the other preserves context-utterance structure for relation representation estimation and interpretation-cue construction. The second stage effectively summarizes complementary local modality evidence and global contextual evidence into the compact final interpretation cue, and the third stage strategically injects this interpretation cue into the final multimodal reasoning stage through a learnable interaction guidance vector. We denote by $H_m^{p}$ the primary modality representations and by $H_m^{s}$ the structure-preserving representations.

\subsection{Context-Utterance Structure Encoding}

Context-Utterance Structure Encoding establishes the structural basis for subsequent interpretation-cue abstraction. It preserves the structural distinction between context and utterance and derives a text-anchored relation representation for the acoustic and visual modalities. We first encode the textual stream with an ALBERT-based textual encoder and derive the text-anchored relation representation. Conditioned on this relation representation, the acoustic and visual streams are then encoded by dual-expert Transformer encoders in the same shared interaction space of dimension $d$. Here, ``nonverbal'' refers specifically to the acoustic and visual modalities. A binary context-utterance partition indicator is used to inject learnable, modality-specific structure embeddings at the input stage, and the same indicator is reused after encoding to separate the hidden sequence into contextual and utterance parts for relation construction. For clarity, we first concisely describe the shared first-stage encoding pipeline without redundant branch superscripts, and formally re-introduce branch-specific notation only when branch-specific outputs are clearly explicitly used.
\par\noindent
\textbf{Text-anchored relation representation.}
Following prior text-guided multimodal modeling studies~\cite{wang2023tetfn,li-li-2025-hne}, we use the textual modality as the anchor for relation modeling. Text provides the most stable semantic reference for estimating the coarse relation between context and the current utterance.
The textual sequence is encoded by the ALBERT-based textual encoder to obtain
$H_t\in\mathbb{R}^{T_t\times d}$, where $T_t=T_c^t+T_u^t$ denotes the length of the joint textual sequence.
Using the same partition indicator, $H_t$ is separated into contextual and utterance parts, and masked mean pooling is applied to obtain a context summary $h_t^{c}$ and an utterance summary $h_t^{u}$. Inspired by previous work~\cite{conneau2017supervised}, we then construct a text-anchored relation representation $r$ by concatenating the context summary, the utterance summary, and their difference.
The difference term explicitly captures the semantic shift from the preceding context to the current utterance.
The relation representation is further mapped by a sigmoid-activated scorer to a stable scalar relation prior score $s\in(0,1)$. It acts as a soft regularizer rather than an exact routing controller. Larger $s$ indicates a stronger inherent consistency tendency between context and utterance, while smaller $s$ indicates a stronger discrepancy tendency.

\par\noindent
\textbf{Relation-guided nonverbal encoding.}
Conditioned on the text-anchored relation representation, the acoustic and visual streams are encoded by relation-guided dual-expert Transformer encoders, where experts correspond to two separate FFN branches. The relation prior score provides a coarse sample-level regularizer, and each modality-specific router makes layer-wise decisions jointly from the current hidden states and the projected relation representation.
For each nonverbal modality $m\in\{a,v\}$, the relation representation $r$ is first projected into the corresponding modality space to obtain a modality-specific relation guidance vector $\tilde r_m$, which is reused across all encoder layers.
At layer $l$, the current hidden states are summarized by sequence-level mean pooling to obtain $z_m^l$.
The concatenation $[z_m^l;\,\tilde r_m]$ is then fed into a lightweight linear router to produce a modality- and layer-specific coefficient $\rho_m^l\in(0,1)$.
The coefficient $\rho_m^l$ controls the trade-off between two parallel feed-forward experts, corresponding to consistency-oriented and discrepancy-oriented nonverbal transformations.
The feed-forward sublayer is thus rewritten as
\begin{equation}
\mathrm{FFN}_{m}^{l}(x)
=
\rho_m^l E_{m,\mathrm{con}}^{l}(x)
+
(1-\rho_m^l) E_{m,\mathrm{dis}}^{l}(x),
\qquad m\in\{a,v\}.
\end{equation}
Accordingly, we define the routing regularization as
\begin{equation}
\mathcal{L}_{\mathrm{gate}}
=
\sum_{m\in\{a,v\}}\sum_l
\mathrm{BCE}(\rho_m^l,\mathrm{sg}(s))
+
\lambda_{\mathrm{bias}}(\tau)\,\mathcal{L}_{\mathrm{bias}},
\end{equation}
where $\mathrm{sg}(\cdot)$ denotes stop-gradient, $\mathcal{L}_{\mathrm{bias}}=\sum_{m\in\{a,v\}}\sum_l\left(\rho_m^l-\frac{1}{2}\right)^2$ is a lightweight balancing regularizer that discourages early routing collapse by keeping the routing coefficients away from premature one-sided concentration, and $\lambda_{\mathrm{bias}}(\tau)$ is linearly decayed with the training epoch $\tau$ so that the balancing effect is stronger in early training and gradually weakened afterward.

\begin{figure}[t]
  \centering
  \includegraphics[width=0.45\textwidth]{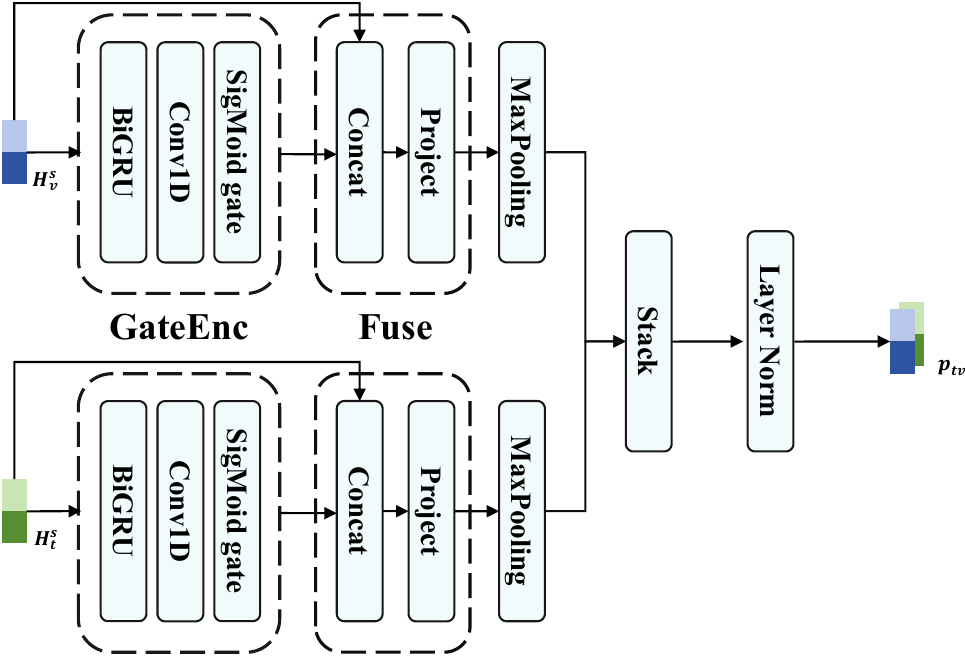}  % 关键修改点
  \caption{Detailed local pairwise cue construction. Here, only the text-visual pair is shown for clarity; the other modality pairs are constructed analogously. \textbf{GateEnc + Fuse + MaxPooling} perform local feature summarization, while the subsequent modules form the \textbf{Pairwise Cue Constructor}.
}  % 保持原始caption
  \label{fig:Local}
\end{figure}

\subsection{Global-Local Interpretation Cue}

Given the structure-preserving representations produced by the first stage, denoted as $H_m^{s} \in \mathbb{R}^{T_m \times d}$ for $m \in \{t,a,v\}$, we next abstract the final interpretation cue for subsequent multimodal reasoning. The local branch preserves utterance-centered local evidence from the joint conversational sequence and explicit pairwise cross-modal organization, while the global branch provides discourse-level interpretation guidance from contextual text.

\par\noindent
\textbf{Local pairwise cue construction.}
The local branch captures utterance-centered local evidence from the joint conversational sequence and pairwise cross-modal compatibility before they are absorbed into the final interpretation cue. It first compresses each modality into a compact local summary and then explicitly constructs modality-pair cues from these summaries.
For each modality $m\in\{t,a,v\}$, the sequence $H_m^s$ is first encoded by a bidirectional GRU, and the recurrent hidden states are further modulated by a lightweight convolutional sigmoid gate to obtain a gated sequence representation:
\begin{equation}
\bar H_m=\mathrm{GateEnc}(H_m^s),
\qquad m\in\{t,a,v\}.
\end{equation}
\begin{equation}
z_m=\mathrm{MaxPool}\,\bigl(\mathrm{Fuse}(H_m^s,\bar H_m)\bigr),
\qquad m\in\{t,a,v\}.
\end{equation}
Here, $\mathrm{GateEnc}(\cdot)$ denotes a BiGRU followed by a lightweight convolutional sigmoid gate for compact local summarization. $\mathrm{Fuse}(\cdot,\cdot)$ denotes concatenation-based gated fusion between the original sequence and the gated sequence, followed by a linear projection.
The three modality summaries are then projected into a shared local space, denoted by $\hat z_m=W_m z_m$, from which three pairwise local cues are constructed as
\begin{equation}
p_{ij}=\operatorname{vec}\!\bigl(\operatorname{Norm}(\operatorname{Stack}(\hat z_i,\hat z_j))\bigr),\quad (i,j)\in\{(t,a),(t,v),(a,v)\},
\end{equation}
where $\operatorname{Stack}(\cdot,\cdot)$ neatly stacks two projected modality summaries along a new pair dimension to form a compact two-slot pair representation, $\operatorname{Norm}(\cdot)$ denotes a lightweight normalization step over the stacked pair representation, and $\operatorname{vec}(\cdot)$ fully vectorizes the normalized pair representation into a local cue vector.

\par\noindent
\textbf{Global evidence readout via textual context query.}
The global branch derives a shared query from the contextual textual states. These contextual textual states serve as the textual anchor for discourse-level interpretation of the current utterance across modalities. Using the structure-preserving textual states $H_t^{s}$ and the same partition indicator as in the first stage, we apply masked mean pooling to the contextual part of the textual sequence to obtain a context summary $h_t^{c}$. This context summary is then projected as the global query, denoted by $\hat q=W_q h_t^{c}$. For each modality $m\in\{t,a,v\}$, the same partition indicator is reused to retain only the utterance-part sequence, denoted by $H_m^{u}$. The global query then reads utterance-level evidence from each modality through cross-attention:
\begin{equation}
g_m=\mathrm{Attn}(\hat q,\, H_m^{u},\, H_m^{u}),
\qquad m\in\{t,a,v\},
\end{equation}
where $\hat q$ serves as the query, $H_m^{u}$ serves as the key and the value.

Finally, the three local pairwise cues and the projected global response are concatenated to form the resulting interpretation cue. Let $\hat g_f=W_o [\, g_t;\, g_a;\, g_v \,]$ denote the projected global response. Then,
\begin{equation}
u_f = [\, p_{ta};\, p_{tv};\, p_{av};\, \hat g_f \,].
\end{equation}

\subsection{Cue-Guided Multimodal Interaction}

Given the modality-specific outputs from the first stage, denoted by
$H_m^{p}\in\mathbb{R}^{T_m\times d}$ for $m\in\{t,a,v\}$,
we initialize the three modality-centered streams as
$H_m^{0}=H_m^{p}$.
Interpretation-Cue-Guided Multimodal Interaction is the final multimodal reasoning stage. Starting from the primary modality representations $H_m^{p}$, it maps the resulting interpretation cue $u_f$ to an interaction guidance vector for subsequent layer-wise multimodal interaction. The interpretation cue is compressed into an interaction guidance vector and injected into all modality-centered streams while preserving their modality-specific sequential structure:
\begin{equation}
G_s=\Gamma(u_f), \qquad G_s\in\mathbb{R}^{1\times d},
\end{equation}
where $\Gamma(\cdot)$ is a guidance projection implemented by one linear layer in the shared interaction space.

\par\noindent
\textbf{Layer-wise guidance-conditioned multimodal reasoning.}
This stage is composed of multiple interaction layers.
At the $l$-th layer, each modality stream is first updated with the interaction guidance vector:
\begin{equation}
\tilde{H}_m^{\,l}
=
H_m^{\,l-1}
+
\Phi_m\!\left(H_m^{\,l-1}, G_s\right),
\qquad m \in \{t,a,v\},
\end{equation}
where $\Phi_m(\cdot,\cdot)$ denotes a dedicated guidance-conditioned attention block consisting of one cross-attention sublayer followed by one elaborate self-attention refinement sublayer, each equipped with standard residual projection and LayerNorm. The current modality stream serves as the query, and the interaction guidance vector provides the conditioning key and value.

Each guidance-aware modality stream then interacts with the other two modalities.
For an anchor modality $m$, let $m_1$ and $m_2$ denote the remaining two modalities.
The two cross-modal responses are computed as
\begin{equation}
R_{m \leftarrow m_1}^{\,l}
=
\mathrm{Attn}\!\left(\tilde{H}_m^{\,l}, \tilde{H}_{m_1}^{\,l}, \tilde{H}_{m_1}^{\,l}\right),
\end{equation}
\begin{equation}
R_{m \leftarrow m_2}^{\,l}
=
\mathrm{Attn}\!\left(\tilde{H}_m^{\,l}, \tilde{H}_{m_2}^{\,l}, \tilde{H}_{m_2}^{\,l}\right).
\end{equation}
Since the informativeness of the two supporting modalities may vary across samples, interaction layers, and feature dimensions, we use an element-wise sigmoid gate to integrate the two cross-modal responses:
\begin{equation}
C_m^{\,l}
=
\beta_m^{\,l} \odot R_{m \leftarrow m_1}^{\,l}
+
\left(1-\beta_m^{\,l}\right) \odot R_{m \leftarrow m_2}^{\,l},
\end{equation}
$\beta_m^{\,l}$ denotes an element-wise sigmoid gate parameterized by affine projections of the two cross-modal responses.
The integrated cross-modal response is then used to update the modality stream:
\begin{equation}
H_m^{\,l}
=
\tilde{H}_m^{\,l}
+
\Psi_m\!\left(C_m^{\,l}\right),
\end{equation}
where $\Psi_m(\cdot)$ denotes a lightweight self-attentive refinement block with residual projection and LayerNorm consistently applied to the rich integrated cross-modal response.

\par\noindent
\textbf{Adaptive multimodal aggregation.}
After the stacked interaction layers, the refined text, acoustic, and visual streams are adaptively aggregated to produce the final multimodal representation. For each modality $m$, we first obtain a temporally pooled modality summary $\bar h_m$ from the final modality stream and then map $\bar h_m$ to a scalar score $o_m$ through a lightweight modality-specific scoring projection. The modality weights are normalized across modalities:
\begin{equation}
\alpha_m
=
\frac{\exp(o_m)}
{\sum\nolimits_{j\in\{t,a,v\}}\exp(o_j)}.
\end{equation}
The final multimodal representation is obtained as
\begin{equation}
z
=
\sum_{m\in\{t,a,v\}}\alpha_m\,\bar{h}_m.
\end{equation}
This representation is finally fed into the classifier for prediction.
\par\noindent
\textbf{Learning Objective.}
The overall training objective combines the task loss and the gate regularization loss:
\begin{equation}
\mathcal{L}
=
\mathcal{L}_{\mathrm{task}}
+
\lambda_{\mathrm{gate}}\,\mathcal{L}_{\mathrm{gate}},
\end{equation}
where $\mathcal{L}_{\mathrm{task}}$ denotes the cross-entropy loss for downstream label classification, $\mathcal{L}_{\mathrm{gate}}$ denotes the gate regularization loss defined in Eq.~(2), and $\lambda_{\mathrm{gate}}$ is a balancing coefficient that controls the contribution of the gate term.
In this way, the model is jointly optimized for downstream prediction and relation-guided expert routing.

\begin{table*}[t]
  \small
  \centering
  \renewcommand{\arraystretch}{1.0}
  \caption{Performance comparison on the MUStARD and MUStARD++ benchmarks. Results for all baselines except DLF and MFMB-Net are taken from the PS2RI paper under its ALBERT-based evaluation protocol. Following the same setting, we further implement DLF and MFMB-Net for a more complete comparison. For our method, the text, audio, and visual encoder settings are kept consistent with PS2RI, except that the original interaction encoder is replaced by our proposed dual-expert design. The publication venue of each baseline is also reported for reference.}
  \label{tab:results}
  \begin{tabular*}{\textwidth}{@{\extracolsep{\fill}} c l c ccc ccc ccc }
    \toprule
    \raisebox{-1.2ex}[0pt][0pt]{\textbf{Benchmark}} &
    \raisebox{-1.2ex}[0pt][0pt]{\textbf{Method}} &
    \raisebox{-1.2ex}[0pt][0pt]{\textbf{Venue}}
      & \multicolumn{3}{c}{\textbf{Entire Set}}
      & \multicolumn{3}{c}{\textbf{Subset 1}}
      & \multicolumn{3}{c}{\textbf{Subset 2}} \\
    \cmidrule(lr){4-12}
    & & & \textbf{Pre(\%)} & \textbf{Rec(\%)} & \textbf{F1(\%)}
      & \textbf{Pre(\%)} & \textbf{Rec(\%)} & \textbf{F1(\%)}
      & \textbf{Pre(\%)} & \textbf{Rec(\%)} & \textbf{F1(\%)} \\
    \midrule
    \multirow{11}{*}{\textbf{MUStARD}}
      & TFN \cite{zadeh2017tensor}     & EMNLP'17 & 49.36 & 45.57 & 48.39 & 53.92 & 49.41 & 50.99 & 45.63 & 47.21 & 46.44 \\
      & LMF \cite{liu2018efficient}    & ACL'18   & 52.14 & 54.41 & 52.66 & 58.45 & 60.78 & 58.57 & 45.46 & 48.04 & 45.58 \\
      & MulT \cite{tsai2019multimodal} & ACL'19   & 51.89 & 55.15 & 52.83 & 54.82 & 58.82 & 56.74 & 44.41 & 50.56 & 48.90 \\
      & MISA \cite{hazarika2020misa} & MM'20 & 54.07 & 49.82 & 52.93 & 60.07 & 55.62 & 58.96 & 43.57 & 48.41 & 46.13 \\
      & MAG \cite{rahman2020integrating} & ACL'20 & 52.76 & 54.27 & 53.85 & 62.53 & 59.74 & 59.16 & 42.56 & 50.42 & 47.32 \\
      & $\mathcal{F}$-MTL \cite{chauhan-etal-2020-sentiment} & ACL'20 & 52.13 & 57.42 & 54.86 & 58.24 & 62.17 & 60.43 & 42.11 & 53.95 & 46.94 \\
      & DMD \cite{Li_2023_CVPR}        & CVPR'23  & 54.29 & 59.08 & 56.27 & 60.71 & 62.49 & 61.71 & 44.97 & 52.82 & 49.66 \\
      & PS2RI \cite{fang2024ps2ri} & MM'24    & 57.46 & 62.06 & 58.45 & 61.73 & 66.47 & 63.52 & 52.42 & 57.65 & 53.50 \\
      & DLF \cite{wang2025dlf}         & AAAI'25  & 52.30 & 63.24 & 55.35 & 57.13 & 68.53 & 62.73 & 52.70 & 55.88 & 52.37 \\
      & MFMB-Net \cite{tao2025mfmb}    & AAAI'25  & 55.24 & 60.29 & 56.66 & 55.82 & 58.82 & 57.08 & 56.48 & 61.76 & 56.71 \\
      & \textbf{CUCI-Net (Ours)}       & --       & \textbf{61.76} & \textbf{67.65} & \textbf{64.37}
                     & \textbf{66.79} & \textbf{70.59} & \textbf{68.63}
                     & \textbf{58.24} & \textbf{64.71} & \textbf{60.28} \\
    \midrule
    \multirow{11}{*}{\textbf{MUStARD++}}
      & TFN \cite{zadeh2017tensor}     & EMNLP'17 & 14.62 & 20.51 & 16.95 &  9.05 & 14.90 & 12.71 & 27.33 & 30.71 & 29.35 \\
      & LMF \cite{liu2018efficient}    & ACL'18   & 15.24 & 19.17 & 17.76 & 13.34 & 15.63 & 14.21 & 26.86 & 23.21 & 24.86 \\
      & MulT \cite{tsai2019multimodal} & ACL'19   & 15.96 & 25.01 & 18.03 & 14.15 & 12.37 & 13.82 & 31.34 & 25.74 & 29.21 \\
      & MISA \cite{hazarika2020misa} & MM'20 & 16.01 & 25.83 & 18.56 & 15.63 & 17.19 & 15.58 & 26.67 & 30.17 & 28.62 \\
      & MAG \cite{rahman2020integrating} & ACL'20 & 17.46 & 22.51 & 19.38 & 17.07 & 20.31 & 18.35 & 26.22 & 35.71 & 29.76 \\
      & $\mathcal{F}$-MTL \cite{chauhan-etal-2020-sentiment} & ACL'20 & 19.29 & 21.33 & 19.82 & 17.61 & 19.06 & 18.41 & 26.26 & 29.77 & 27.58 \\
      & DMD \cite{Li_2023_CVPR}        & CVPR'23  & 20.37 & 24.16 & 22.81 & 22.14 & 17.31 & 19.04 & 34.56 & 33.46 & 34.07 \\
      & PS2RI \cite{fang2024ps2ri} & MM'24    & 24.06 & 25.83 & 24.28 & 22.96 & 18.75 & 19.27 & 37.59 & 33.93 & 35.04 \\
      & DLF \cite{wang2025dlf}         & AAAI'25  & 23.18 & 26.47 & 23.56 & 22.41 & 19.52 & 19.74 & 35.86 & 34.71 & 34.28 \\
      & MFMB-Net \cite{tao2025mfmb}    & AAAI'25  & 24.02 & 25.94 & 23.97 & 21.94 & 18.87 & 19.02 & 37.21 & 35.36 & 36.08 \\
      & \textbf{CUCI-Net (Ours)}       & --       & \textbf{27.43} & \textbf{29.89} & \textbf{28.50}
                     & \textbf{25.74} & \textbf{23.27} & \textbf{25.79}
                     & \textbf{41.13} & \textbf{36.82} & \textbf{39.17} \\
    \bottomrule
  \end{tabular*}
\end{table*}

\section{Experiment}

\subsection{Dataset}
\par\noindent
\textbf{MUStARD.}
MUStARD is a multimodal conversational sarcasm detection dataset proposed by Castro \textit{et al.}~\cite{castro-etal-2019-towards}. It consists of textual, visual, and acoustic modalities, and contains 690 \textit{dialogue instances}, where each instance includes a contextual segment and a current utterance annotated with a sarcasm label. Since sarcasm in MUStARD is often determined by the dependency between the preceding dialogue and the current utterance, the dataset is widely used to evaluate context-sensitive multimodal understanding. This setting requires models to jointly capture contextual cues, utterance-level semantics, and nonverbal signals, rather than relying only on the isolated target utterance. Later, Chauhan \textit{et al.}~\cite{chauhan-etal-2020-sentiment} further extended MUStARD with explicit and implicit sentiment annotations.
\par\noindent
\textbf{MUStARD++.}
MUStARD++ is an extension of MUStARD built by Ray \textit{et al.}~\cite{ray-etal-2022-multimodal}. It retains textual, visual, and acoustic modalities, expands the dataset to 1202 \textit{dialogue instances}, and provides richer annotations including sarcasm, sentiment, emotion, valence, arousal, and sarcasm type. Compared with MUStARD, MUStARD++ introduces broader annotation coverage and greater diversity in contextual and affective patterns, making it a more challenging benchmark for conversational multimodal reasoning. Its multi-level affective labels further allow evaluation beyond binary sarcasm detection, especially on whether models can distinguish different emotional and pragmatic effects under dialogue context.
\par\noindent

\subsection{Implementation Details}
\textbf{Optimization settings.}
Following prior work~\cite{hasan2021hkt,fang2024ps2ri}, we adopt modality-specific learning rates for different encoders and optimization blocks. Specifically, the learning rate is set to $3\times10^{-3}$ for the acoustic and visual encoders, and $2\times10^{-6}$ for the language encoder and the remaining modules. We use Adam with cosine learning rate decay. The dropout rate is set to 0.4, and early stopping with a patience of 10 is used to alleviate overfitting.

\par\noindent
\textbf{Model configuration.}
For Context-Utterance Structure Encoding, we use a 12-layer ALBERT for text, an 8-layer Transformer for vision, and a 1-layer Transformer for acoustics, with a unified hidden dimension of 192. The framework contains two architecture-identical but parameter-independent branches: a primary branch for final multimodal reasoning and a structure-preserving branch for interpretation-cue construction. For relation-guided nonverbal encoding, the two experts are initialized by duplicating the standard feed-forward block and then optimized separately. The gate regularization coefficient is set to $\lambda_{\mathrm{gate}}=0.05$.

\subsection{Main Result}

\par\noindent
\textbf{Task Settings.}
We evaluate CUCI-Net on context-dependent affect recognition using MUStARD and MUStARD++.
As shown in Table~\ref{tab:results}, the model predicts the \emph{utterance sentiment polarity} label on MUStARD and the \emph{utterance implicit emotion} label on MUStARD++.
For both datasets, we further report results on two sarcasm-conditioned subsets:
Subset~1 denotes samples with $\mathrm{sar}=1$, while Subset~2 denotes samples with $\mathrm{sar}=0$.
This split examines whether the model can handle both sarcastic cases with context-shifted affective meanings and non-sarcastic cases with more direct surface expressions.
Precision, Recall, and F1 are reported as evaluation metrics.

\par\noindent
\textbf{Baselines.}
We compare CUCI-Net with representative multimodal baselines in Table~\ref{tab:results}, including TFN, LMF, MulT, MISA, MAG, F-MTL, DMD, PS2RI, DLF, and MFMB-Net~\cite{zadeh2017tensor,liu2018efficient,tsai2019multimodal,hazarika-etal-2018-conversational,rahman2020integrating,chauhan-etal-2020-sentiment,Li_2023_CVPR,fang2024ps2ri,wang2025dlf,tao2025mfmb}.
These methods cover tensor-based fusion, low-rank fusion, cross-modal attention, language-guided fusion, representation disentanglement, and recent relation-aware multimodal modeling.
This comparison verifies whether explicit context--utterance structure modeling and interpretation cue construction bring advantages over holistic fusion and implicit contextual encoding.

\begin{table*}[t]
\centering
\small
\renewcommand{\arraystretch}{1.10}
\setlength{\tabcolsep}{5pt}

\begin{minipage}[t]{0.485\textwidth}
  \centering
  \caption{Ablation results on Context-Utterance Structure Encoding. F1 (\%) is reported on MUStARD and MUStARD++ over the Entire Set and Subset 1.}
  \label{tab:ablation_structure}
  \begin{tabular*}{\linewidth}{@{}l@{\extracolsep{\fill}}cccc@{}}
    \toprule
    \multirow{2}{*}{\textbf{Variant}} 
    & \multicolumn{2}{c}{\textbf{MUStARD}} 
    & \multicolumn{2}{c}{\textbf{MUStARD++}} \\
    \cmidrule(lr){2-3}\cmidrule(lr){4-5}
    & \textbf{Entire} & \textbf{S1} & \textbf{Entire} & \textbf{S1} \\
    \midrule
    Full model                    & \textbf{64.37} & \textbf{68.63} & \textbf{28.50} & \textbf{25.79} \\
    w/o role embeddings           & 60.06 & 61.64 & 26.52 & 25.13 \\
    w/o dual-experts              & 56.78 & 61.23 & 23.92 & 21.16 \\
    w/o independent dual branches & 47.79 & 45.23 & 13.29 & 11.72 \\
    \bottomrule
  \end{tabular*}
\end{minipage}
\hspace{0.02\textwidth}
\begin{minipage}[t]{0.485\textwidth}
  \centering
  \caption{Ablation results on pairwise local cue combinations. F1 (\%) is reported on MUStARD and MUStARD++ over the Entire Set and Subset 1.}
  \label{tab:ablation_cue_pairs}
  \begin{tabular*}{\linewidth}{@{}l@{\extracolsep{\fill}}cccc@{}}
    \toprule
    \multirow{2}{*}{\textbf{Variant}} 
    & \multicolumn{2}{c}{\textbf{MUStARD}} 
    & \multicolumn{2}{c}{\textbf{MUStARD++}} \\
    \cmidrule(lr){2-3}\cmidrule(lr){4-5}
    & \textbf{Entire} & \textbf{S1} & \textbf{Entire} & \textbf{S1} \\
    \midrule
    $(t,a)+(t,v)+(a,v)$ & \textbf{64.37} & \textbf{68.63} & \textbf{28.50} & \textbf{25.79} \\
    $(t,v)+(t,a)$       & 61.92 & 67.35 & 26.89 & 24.09 \\
    $(t,a)+(a,v)$       & 62.00 & 66.78 & 25.91 & 24.37 \\
    $(t,v)+(a,v)$       & 58.82 & 60.28 & 22.28 & 23.02 \\
    \bottomrule
  \end{tabular*}
\end{minipage}

\vspace{2.0em}

\begin{minipage}[t]{0.485\textwidth}
  \centering
  \caption{Ablation results on Global-Local Interpretation Cue Construction. F1 (\%) is reported on MUStARD and MUStARD++ over the Entire Set and Subset 1.}
  \label{tab:ablation_cue}
  \begin{tabular*}{\linewidth}{@{}l@{\extracolsep{\fill}}cccc@{}}
    \toprule
    \multirow{2}{*}{\textbf{Variant}} 
    & \multicolumn{2}{c}{\textbf{MUStARD}} 
    & \multicolumn{2}{c}{\textbf{MUStARD++}} \\
    \cmidrule(lr){2-3}\cmidrule(lr){4-5}
    & \textbf{Entire} & \textbf{S1} & \textbf{Entire} & \textbf{S1} \\
    \midrule
    Full model            & \textbf{64.37} & \textbf{68.63} & \textbf{28.50} & \textbf{25.79} \\
    w/o local cue branch  & 58.82 & 56.71 & 24.63 & 21.29 \\
    w/o global cue branch & 61.27 & 63.35 & 25.21 & 23.61 \\
    \bottomrule
  \end{tabular*}
\end{minipage}
\hspace{0.02\textwidth}
\begin{minipage}[t]{0.485\textwidth}
  \centering
  \caption{Ablation results on Interpretation-Cue-Guided Multimodal Interaction. F1 (\%) is reported on MUStARD and MUStARD++ over the Entire Set and Subset 1.}
  \label{tab:ablation_interaction}
  \begin{tabular*}{\linewidth}{@{}l@{\extracolsep{\fill}}cccc@{}}
    \toprule
    \multirow{2}{*}{\textbf{Variant}} 
    & \multicolumn{2}{c}{\textbf{MUStARD}} 
    & \multicolumn{2}{c}{\textbf{MUStARD++}} \\
    \cmidrule(lr){2-3}\cmidrule(lr){4-5}
    & \textbf{Entire} & \textbf{S1} & \textbf{Entire} & \textbf{S1} \\
    \midrule
    Full model               & \textbf{64.37} & \textbf{68.63} & \textbf{28.50} & \textbf{25.79} \\
    w/o guidance cue         & 60.06 & 58.24 & 26.89 & 24.09 \\
    w/o adaptive aggregation & 58.82 & 57.68 & 25.67 & 22.32 \\
    \bottomrule
  \end{tabular*}
\end{minipage}
\end{table*}

\par\noindent
\textbf{Performance Analysis.}
As shown in Table~\ref{tab:results}, CUCI-Net achieves the best overall performance on both MUStARD and MUStARD++.
On the Entire Set, CUCI-Net consistently outperforms all baselines in Precision, Recall, and F1, demonstrating its effectiveness for context-dependent affect recognition.
The gains are also maintained on both sarcasm-conditioned subsets, indicating that CUCI-Net handles context-shifted sarcastic cases as well as non-sarcastic cases with more direct affective expressions.
Overall, these results show the benefit of preserving context--utterance structure and using interpretation cues to guide multimodal interaction.

\subsection{Ablation Study}
We conduct comprehensive ablation studies to systematically examine the contributions of the major designs in CUCI-Net, including the Context-Utterance Structure Encoding stage, the Global-Local Interpretation Cue Construction stage, and the Interpretation-Cue-Guided Multimodal Interaction stage. In particular, we investigate the effects of the text-anchored relation-guided dual-expert encoding, the global-local interpretation cue construction, and the cue-guided multimodal interaction mechanism, in order to clarify how each component contributes to the overall conversational multimodal understanding performance.

\par\noindent
\textbf{Ablation on Context-Utterance Structure Encoding.}
Table~\ref{tab:ablation_structure} evaluates the first stage of CUCI-Net.
The first-stage ablation mainly reveals what kind of representation the later cue module actually needs. When role embeddings are removed, the model loses an explicit marker of whether a token belongs to context or to the current utterance, so the projected sequence becomes less suitable for relation construction; this effect is especially visible on Subset~1 of MUStARD, where fine-grained context-utterance distinction matters more. Replacing the dual-expert design with a single path further weakens performance, suggesting that nonverbal encoding should preserve multiple transformation tendencies before cue construction rather than compressing them into one undifferentiated stream. The most severe degradation appears when the two branches are merged, which shows that the primary reasoning branch and the structure-preserving branch should serve different purposes: one for downstream multimodal decision making, and the other for maintaining a cleaner structural view for interpretation-cue derivation.

\par\noindent
\textbf{Ablation on Global-Local Interpretation Cue Construction.}
Table~\ref{tab:ablation_cue} evaluates the second stage of CUCI-Net.
The second-stage ablation clarifies how the interpretation cue should be composed rather than merely confirming that “more branches are better.” Removing either the local cue branch or the global cue branch weakens performance on both datasets, but the larger drop without the local branch suggests that discourse-level context is most useful when it is grounded in concrete pairwise evidence from the current utterance. Table~\ref{tab:ablation_cue_pairs} further shows that the full three-pair setting $(t,v)+(t,a)+(a,v)$ performs best, indicating that the cue should summarize multiple complementary local relations instead of relying on one dominant pair. The pairwise comparison also suggests different roles for the three relations: the two combinations containing $(t,a)$ remain relatively competitive, implying that text-audio interaction provides a comparatively stable semantic anchor for contextual incongruity, whereas the setting without $(t,a)$, i.e., $(t,v)+(a,v)$, degrades the most, showing that visual-acoustic evidence alone is not sufficient to support local interpretation.

\par\noindent
\textbf{Ablation on Interpretation-Cue-Guided Multimodal Interaction.}
Table~\ref{tab:ablation_interaction} evaluates the third stage of CUCI-Net.
The third-stage ablation highlights the difference between having an interpretation cue and actually exploiting it effectively. Removing the guidance cue turns the interaction stack into a largely unguided multimodal encoder, which weakens the model because the later layers no longer receive an explicit relational condition derived from context and utterance. Removing adaptive aggregation causes a further drop, suggesting that even after cue-guided interaction, the resulting modality summaries should not be treated as uniformly informative. Instead, the final decision still depends on a sample-specific consolidation step that can emphasize the most reliable modality mixture for the current conversational case.

\section{Further Study}

\subsection{Layer Sensitivity Analysis}

We further examine how interaction depth affects CUCI-Net.
As the number of interaction layers increases, performance first improves from shallow to moderate depths and then declines when the stack becomes too deep, indicating a transition from insufficient interaction to over-transformation.
The initial improvement suggests that a small number of layers cannot adequately integrate context-utterance dependency with cross-modal evidence, whereas moderate depth enables structured multimodal interactions.
This also implies that the interpretation cue needs propagation steps to guide multimodal representation learning.

This trend is consistently observed on both datasets and across the Entire Set, Subset~1, and Subset~2, suggesting that the effect of layer depth reflects a general property of the interaction module rather than a phenomenon restricted to one partition.
In particular, the peak performance at moderate depth indicates that CUCI-Net does not benefit from unlimited interaction stacking; instead, once the key contextual relations have been sufficiently propagated, additional layers may introduce redundant transformations and weaken the discriminative structure formed in earlier stages.
Therefore, a moderate interaction depth provides a better balance between cue-guided reasoning and representation stability.

\begin{figure}[t]
    \centering
    \includegraphics[width=1\linewidth]{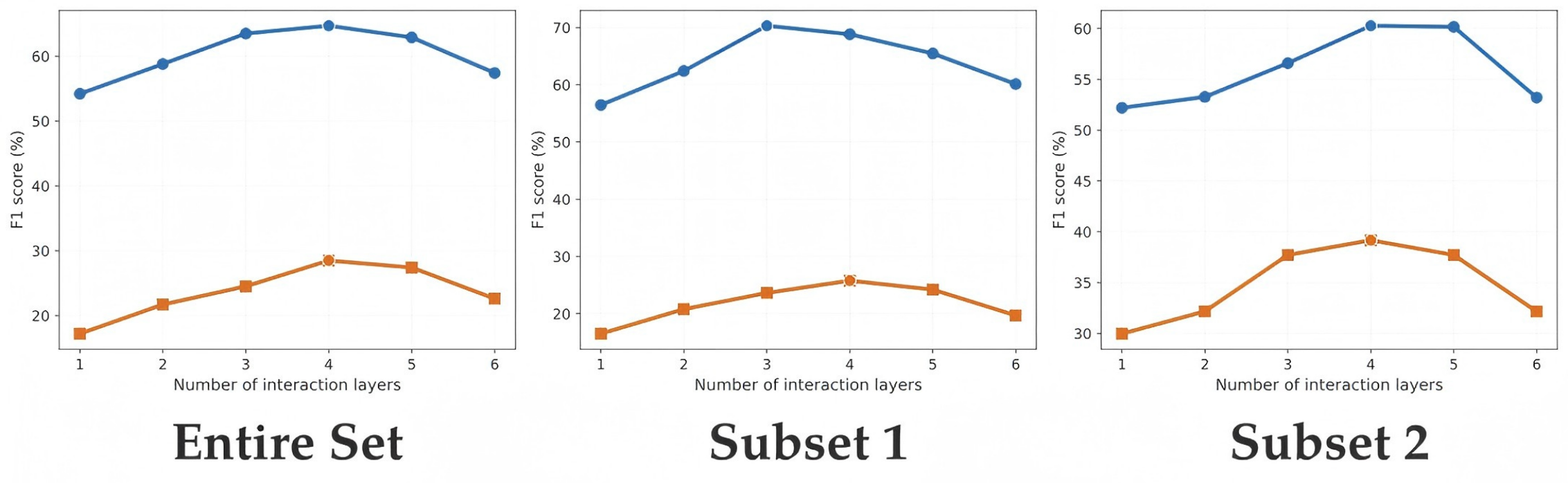}
    \caption{Layer sensitivity analysis of CUCI-Net on MUStARD and MUStARD++. We report F1 (\%) on the Entire Set, Subset~1, and Subset~2 under different interaction depths. Blue and orange curves correspond to MUStARD and MUStARD++, respectively.}
    \label{fig:layerwise_performance}
    \vspace{-0.2cm}
\end{figure}

\subsection{Routing Behavior Analysis}

To further examine whether the proposed relation-guided routing learns meaningful weight allocation patterns, we visualize the routing weights of the acoustic and visual branches.
Since the acoustic branch contains only one Transformer layer, we directly report its routing matrix.
For the visual branch, we visualize the fifth layer, which lies in the stable high-alignment region indicated by the layer-wise analysis and therefore provides a representative view of mature routing behavior.

\begin{figure}[t]
    \centering
    \includegraphics[width=1\linewidth]{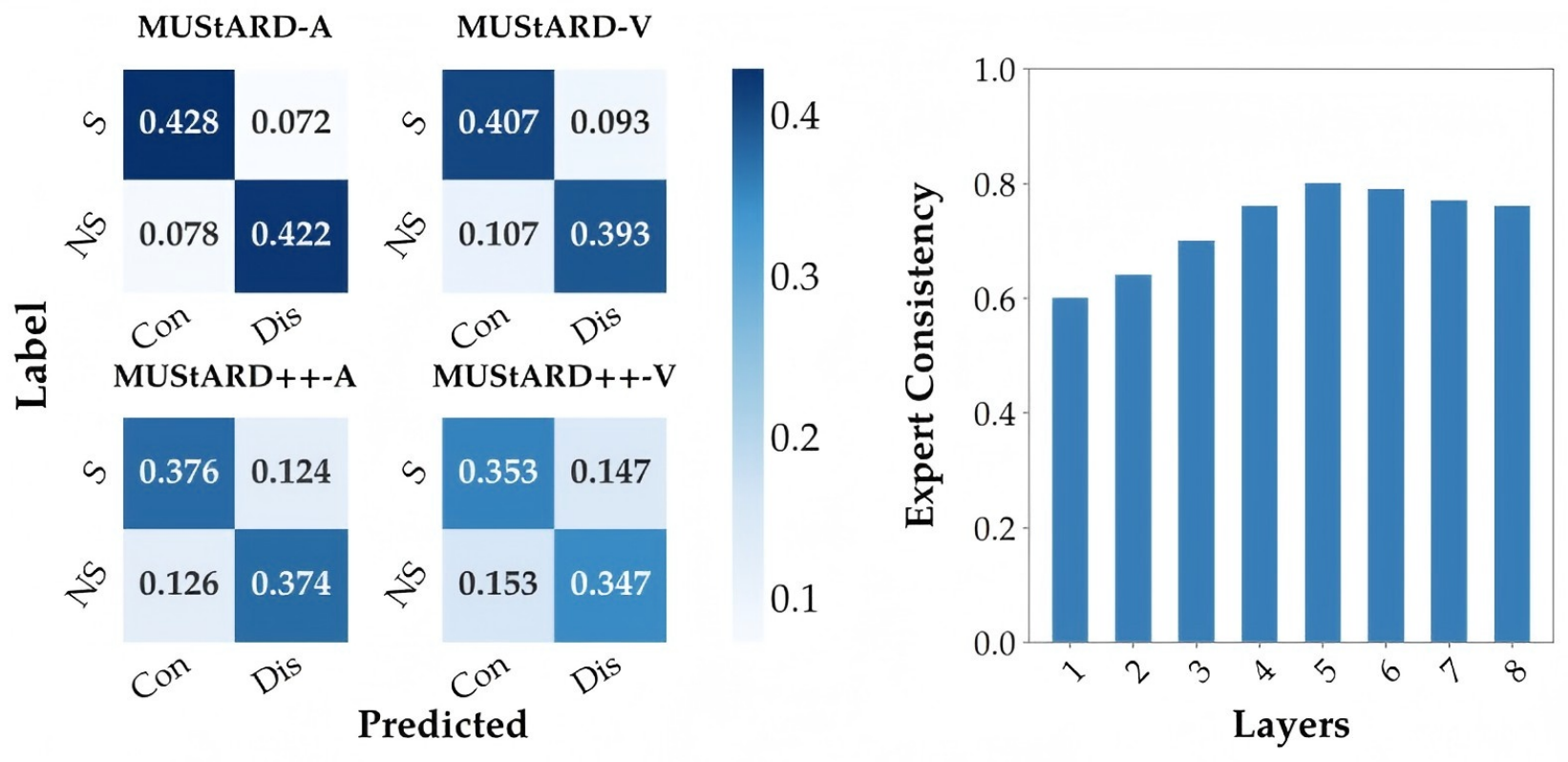}
    \caption{Left: routing heatmaps of the acoustic branch and the fifth visual layer on MUStARD and MUStARD++, illustrating the label-dependent expert preferences, where S/NS denote sarcastic and non-sarcastic subsets, and Con/Dis denote the consistency- and discrepancy-oriented experts. Right: layer-wise expert consistency score of the visual branch on MUStARD, showing how expert specialization becomes clearer across depth.}
    \label{fig:routing_performance}
    \vspace{-0.2cm}
\end{figure}

As shown in Fig.~\ref{fig:routing_performance}, both the acoustic and visual branches exhibit clear label-dependent weight allocation patterns on MUStARD and MUStARD++.
Each entry in the routing matrix denotes the average routing weight assigned to a specific expert under a given label subset.
The diagonal entries consistently dominate the off-diagonal ones, indicating that sarcastic and non-sarcastic samples tend to allocate larger weights to different experts.
This suggests that the routing module captures subset-sensitive relation patterns rather than assigning experts uniformly.
This trend is more pronounced on MUStARD, whereas the routing matrices on MUStARD++ remain structured but comparatively less separable, consistent with its greater diversity and difficulty.
The observation suggests that the proposed routing mechanism does not collapse to uniform allocation, but instead learns subset-sensitive expert preferences that remain interpretable across datasets.

To further quantify how this weight allocation evolves across depth, we compute a layer-wise expert consistency score for the visual branch on MUStARD by averaging the two diagonal entries of the routing matrix at each layer.
This score reflects the average routing weight assigned to the expert that is more aligned with the corresponding subset.
The resulting bar plot shows that the expert consistency score increases from shallow layers to middle layers and then remains in a relatively stable high-value region.
This indicates that expert specialization is progressively refined through the hierarchy rather than being imposed at the first layer.
It also motivates our choice of the fifth visual layer for visualization, since by this stage routing weights have become clearly structured while preserving enough variation to reveal the specialization process.

\section{Conclusion}
We proposed CUCI-Net, a three-stage interpretation-cue-guided framework for conversational multimodal understanding. CUCI-Net preserves the context-utterance structure during encoding, abstracts the resulting dependency into an explicit interpretation cue, and injects this cue into the final multimodal reasoning stage for context-conditioned prediction. Extensive experiments on MUStARD, MUStARD++, and UR-FUNNY demonstrate that the proposed framework achieves competitive strong performance under the current comparison setting, while the ablation studies further verify the core effectiveness of its three major stages, and in-depth analyses further illustrate the rationality and interpretability of our cue-guided reasoning mechanism.

\appendix

\section{Implementation Details}

\textbf{Feature Extraction.}
Following the standard preprocessing protocol for conversational multimodal understanding, we extract textual, acoustic, and visual features for each conversational sample and organize them into a unified input format for subsequent modeling.
Each sample consists of the preceding conversational context and the target utterance.
For text, we use ALBERT~\cite{lan2020albert} to obtain 768-dimensional token-level representations on all three benchmarks.
For the nonverbal modalities, we directly adopt the released benchmark features for fair comparison with prior work.
Specifically, on MUStARD, we use 81-dimensional COVAREP~\cite{degottex2014covarep} acoustic features and 91-dimensional OpenFace 2.0~\cite{baltrusaitis2018openface} visual features.
On MUStARD++, we use the released 513-dimensional acoustic features and 768-dimensional visual features.
On UR-FUNNY, we use 81-dimensional COVAREP acoustic features and 371-dimensional OpenFace 2.0 visual features.

\par\noindent
\textbf{Input Construction.}
For each sample, we construct a paired context--utterance sequence.
The preceding utterances are concatenated into a context sequence, while the current utterance is treated as the target utterance.
After ALBERT tokenization, the textual input is organized as
\begin{equation}
X^{t} = [\text{[CLS]}; C^{t}; \text{[SEP]}; U^{t}; \text{[SEP]}],
\end{equation}
where $C^{t}$ and $U^{t}$ denote the subword token sequences of the context and the target utterance, respectively.

Since the acoustic and visual features are originally extracted at the word level, they are aligned to the subword timeline through a word-to-subword copy operation, where each subword token inherits the acoustic or visual feature vector of its source word.
Let $\tilde{C}^{m}$ and $\tilde{U}^{m}$, where $m \in \{a,v\}$, denote the resulting subword-aligned context and target sequences.
To keep all modalities synchronized with the text input, zero vectors are inserted at the special-token positions, yielding
\begin{equation}
X^{m} = [\mathbf{0}; \tilde{C}^{m}; \mathbf{0}; \tilde{U}^{m}; \mathbf{0}], \quad m \in \{a,v\},
\end{equation}
where the three zero vectors correspond to \text{[CLS]}, the first \text{[SEP]}, and the final \text{[SEP]}, respectively.

Finally, all modality sequences are padded to the same maximum length.
We further assign segment IDs to distinguish the two parts of the conversation: segment ID 0 is used for \text{[CLS]} + context + \text{[SEP]}, while segment ID 1 is used for target utterance + \text{[SEP]}.
In this way, the model receives temporally aligned multimodal inputs while explicitly preserving the boundary between contextual evidence and the target utterance.

\section{Cross-Dataset Experiment}
\subsection{Cross-Dataset Sentiment Generalization}
\label{subsec:cross_dataset_sentiment}

\par\noindent
\textbf{Task Setting.}
We further evaluate CUCI-Net on CMU-MOSEI and CMU-MOSI to examine whether the proposed architecture can generalize to standard utterance-level multimodal sentiment benchmarks.
Following the official data splits, we adopt the standard non-neutral setting, where neutral samples are removed and only non-neutral samples are retained for evaluation.
We report F1, MAE, and Acc-2 as the primary metrics.

\par\noindent
\textbf{Dataset Adaptation.}
Since CMU-MOSEI and CMU-MOSI do not provide explicit conversational context in the same form as MUStARD-style benchmarks, we use a simple input adaptation for CUCI-Net.
Specifically, for each sample, we duplicate the three-modal utterance inputs and use the duplicated copy as a pseudo-context branch, the original sample is treated as the target utterance branch.
This design does not introduce additional external information, but preserves the context--utterance input format required by CUCI-Net.

\par\noindent
\textbf{Baselines.}
As shown in Table~\ref{tab:mosi_mosei_main}, we compare CUCI-Net with representative multimodal sentiment analysis baselines, including MulT, MISA, DMD, DLF, and MFMB-Net.
For fair comparison, most baseline results are quoted from the DLF paper, except for MFMB-Net, whose results are taken from its original paper.

\par\noindent
\textbf{Analysis.}
CUCI-Net achieves the best F1 and Acc-2 on both CMU-MOSEI and CMU-MOSI, indicating that its structure-aware representation learning, interpretation cue construction, and cue-guided multimodal interaction are not restricted to context-dependent sarcasm-style benchmarks.
Meanwhile, its MAE is not always the best, which is reasonable because CUCI-Net is mainly developed for discriminative context-sensitive understanding rather than fine-grained sentiment regression.
Overall, these results provide auxiliary evidence that the proposed architecture can transfer to broader utterance-level multimodal sentiment analysis settings.

\begin{table}[t]
  \centering
  \caption{Performance comparison on CMU-MOSEI and CMU-MOSI under the non-neutral setting. For fair comparison, the baseline results are quoted from the DLF paper except for MFMB-Net, whose results are taken from its original paper.}
  \label{tab:mosi_mosei_main}
  \small
  \setlength{\tabcolsep}{4pt}
  \renewcommand{\arraystretch}{1.05}
  \resizebox{\columnwidth}{!}{%
  \begin{tabular}{lccccccc}
    \toprule
    \textbf{Method} & \textbf{Venue} &
    \multicolumn{3}{c}{\textbf{CMU-MOSEI}} &
    \multicolumn{3}{c}{\textbf{CMU-MOSI}} \\
    \cmidrule{3-5}\cmidrule{6-8}
    & &
    \textbf{F1} & \textbf{MAE} & \textbf{Acc-2} &
    \textbf{F1} & \textbf{MAE} & \textbf{Acc-2} \\
    \midrule
    MulT\cite{tsai-etal-2019-multimodal}   & ACL'19  & 82.30 & 0.580 & 82.50 & 82.00 & 0.871 & 83.00 \\
    MISA\cite{hazarika2020misa}            & MM'20   & 84.66 & 0.558 & 84.67 & 83.58 & 0.777 & 83.54 \\
    DMD\cite{li2023dmd}                    & CVPR'23 & 84.62 & 0.543 & 84.62 & 83.29 & 0.752 & 83.23 \\
    DLF\cite{wang2025dlf}                  & AAAI'25 & 85.27 & \textbf{0.530} & 85.42 & 85.04 & 0.731 & 85.06 \\
    MFMB-Net\cite{tao2025mfmb}             & AAAI'25 & 85.10 & 0.532 & 85.10 & 86.00 & \textbf{0.709} & 85.70 \\
    \midrule
    \textbf{CUCI-Net (Ours)}               & --      & \textbf{86.32} & 0.537 & \textbf{86.26} & \textbf{87.00} & 0.749 & \textbf{86.20} \\
    \bottomrule
  \end{tabular}%
  }
\end{table}

\begin{table}[t]
    \centering
    \small
    \renewcommand{\arraystretch}{1.05}
    \setlength{\tabcolsep}{5pt}
    \caption{Results on UR-FUNNY and MUStARD are reported following the standard evaluation protocols and benchmark settings from the original publications of baseline methods.}
    \label{tab:glomo_compare}
    \begin{tabular}{lccc}
        \toprule
        \textbf{Method} & \textbf{Venue} & \textbf{UR-FUNNY} ($\uparrow$) & \textbf{MUStARD} ($\uparrow$) \\
        \midrule
        MISA~\cite{hazarika2020misa}      & MM'20    & 69.82 & 66.18 \\
        MAG~\cite{rahman2020integrating}      & ACL'20   & 67.20 & 69.12 \\
        HKT~\cite{hasan2021hkt}           & AAAI'21  & 77.36 & 79.41 \\
        MCL~\cite{mai2023mcl}             & IF'23      & 77.70 & 77.90 \\
        MGCL~\cite{mai2023mgcl}           & IF'23       & 78.10 & 77.90 \\
        \midrule
        \textbf{CUCI-Net (Ours)}            & --       & \textbf{78.92} & \textbf{80.21} \\
        \bottomrule
    \end{tabular}
\end{table}

\subsection{Non-Literal Expression}
\label{subsec:humor_sarcasm_generalization}

\par\noindent
\textbf{Task Setting.}
We also evaluate CUCI-Net on UR-FUNNY and MUStARD to further test its generalization under binary humor and sarcasm recognition settings.
Following prior work, the model predicts the humor label on UR-FUNNY and the sarcasm label on MUStARD.
Acc-2 is reported for both datasets under their standard evaluation protocols.

\par\noindent
\textbf{Datasets.}
UR-FUNNY is a multimodal humor detection dataset proposed by Hasan \textit{et al.}~\cite{hasan-etal-2019-ur-funny}, containing 16{,}514 utterance instances from 1{,}866 TED talks with textual, acoustic, and visual modalities.
MUStARD is a multimodal conversational sarcasm detection dataset proposed by Castro \textit{et al.}~\cite{castro-etal-2019-towards}, containing 690 dialogue instances with contextual segments and target utterances annotated with sarcasm labels.
Compared with utterance-level sentiment analysis, these two datasets require stronger modeling of non-literal intent, humor cues, and context-sensitive interpretation.

\par\noindent
\textbf{Baselines.}
As shown in Table~\ref{tab:glomo_compare}, we compare CUCI-Net with representative multimodal baselines, including MISA, MAG, HKT, MCL, and MGCL.
baseline results are reported from their publications.

\par\noindent
\textbf{Analysis.}
CUCI-Net achieves the best performance on both UR-FUNNY and MUStARD, obtaining 78.92 on UR-FUNNY and 80.21 on MUStARD.
The improvement over recent baselines suggests that CUCI-Net can handle not only affect polarity prediction, but also broader pragmatic understanding problems such as humor and sarcasm recognition.
This result is consistent with our motivation: by preserving context--utterance structure and converting their dependency into an explicit interpretation cue, the model can better support multimodal reasoning when the final label depends on contextual incongruity, non-literal intent, or cross-modal evidence rather than isolated utterance semantics alone.

\begin{algorithm}[t]
\caption{Overall Forward Procedure of CUCI-Net}
\label{alg:cucinet}
\begin{algorithmic}[1]
\Require Multimodal inputs $\{C^m,U^m\}_{m\in\{t,a,v\}}$
\Ensure Prediction $\hat{y}$

\For{each modality $m \in \{t,a,v\}$}
    \State Construct the joint sequence $X^m = [C^m; U^m]$
\EndFor

\Statex
\State \textbf{Stage 1: Context--Utterance Structure Encoding}
\For{each modality $m \in \{t,a,v\}$}
    \State Encode $X^m$ into primary features $H_m^p$ and structure-preserving features $H_m^s$
\EndFor
\State Derive the text-anchored relation representation $r$ from $H_t^s$

\Statex
\State \textbf{Stage 2: Global--Local Interpretation Cue Construction}
\State Summarize modality-specific local evidence from $\{H_m^s\}_{m\in\{t,a,v\}}$
\State Construct pairwise local cues and global evidence guided by the textual context
\State Form the interpretation cue $u_f$ and map it to the guidance token $G_s=\Gamma(u_f)$

\Statex
\State \textbf{Stage 3: Interpretation-Cue-Guided Multimodal Interaction}
\State Initialize $H_m^{p,0}=H_m^p$ for each modality $m$
\For{$l=1$ to $L$}
    \For{each modality $m \in \{t,a,v\}$}
        \State Update $H_m^{p,l}=\Phi_m(H_m^{p,l-1},G_s)$
    \EndFor
\EndFor

\State Fuse $\{H_m^{p,L}\}_{m\in\{t,a,v\}}$ into $Z$
\State Predict $\hat{y}=\mathrm{Classifier}(Z)$
\State \Return $\hat{y}$
\end{algorithmic}
\end{algorithm}

\section{Visualization}

\par
\begin{figure}[t]
    \centering
    \includegraphics[width=\linewidth]{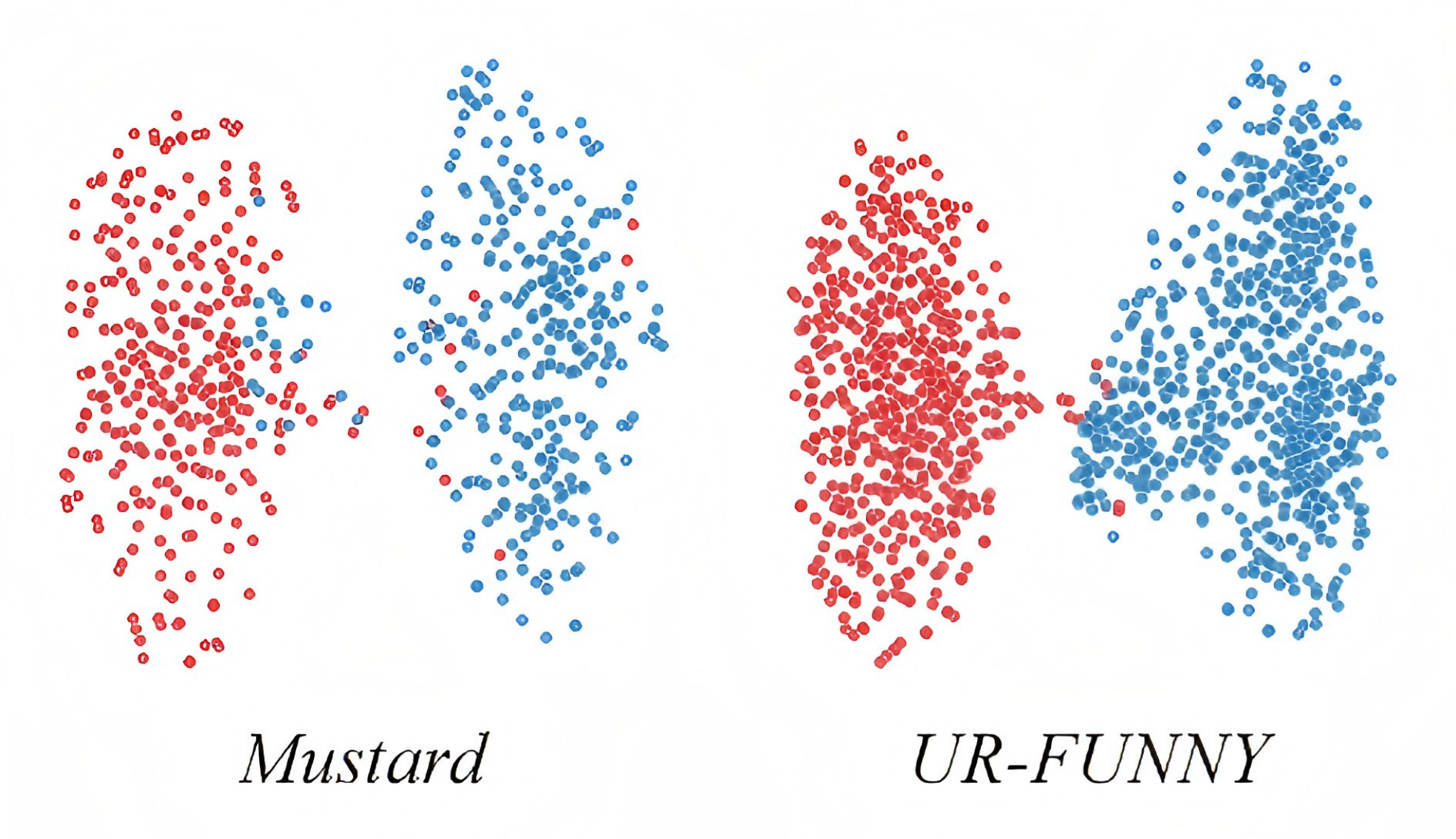}
    \caption{\textbf{t-SNE visualization of the learned feature distributions.} Red points denote sarcastic/funny samples, and blue points denote non-sarcastic/non-funny samples. Left: MUStARD. Right: UR-FUNNY.}
    \label{fig:tsne}
\end{figure}

\noindent
\textbf{Feature Visualization.}
Fig.~\ref{fig:tsne} presents the t-SNE visualization of the learned representations on MUStARD and UR-FUNNY.
On both datasets, samples from the two classes show a clear tendency to form distinguishable distributions, suggesting that CUCI-Net learns discriminative multimodal representations for context-dependent understanding.
On MUStARD, the two classes are largely separable, although a small overlap remains near the decision boundary, which is reasonable given the ambiguity of sarcasm recognition.
On UR-FUNNY, the class structure appears more compact and better separated, indicating that the learned representations also transfer well to humor recognition.
Overall, these visualizations provide strong qualitative evidence that CUCI-Net can effectively capture class-relevant multimodal patterns and further organize them into a more discriminative feature space.

\begin{table}[t]
\centering
\caption{Representative case studies of CUCI-Net on MUStARD. Both cases are correctly classified by CUCI-Net and illustrate two typical forms of context-dependent multimodal understanding. For each case, we report the sample ID, the SI and SE labels, the preceding conversational context, and the target utterance to show how the final meaning depends on context--utterance interaction rather than on the target utterance alone.}
\label{tab:case_study}
\footnotesize
\setlength{\tabcolsep}{3pt}
\renewcommand{\arraystretch}{1.08}
\begin{tabularx}{\columnwidth}{p{0.8cm}p{0.6cm}p{0.6cm}X X}
\toprule
\textbf{hid} & \textbf{SI} & \textbf{SE} & \textbf{Context} & \textbf{Utterance} \\
\midrule
1\_3293 & Neg & Pos &
``Are you interested in physics?''\newline
``Oh, I find it fascinating.''\newline
``If I hadn't gone into microbiology, I probably would have gone into ice dancing.''\newline
``Actually, my tests of the Bohm quantum interference effect have reached an interesting point.''\newline
``Right now we're testing the phase shift due to an electric potential.''\newline
``That's amazing.''\newline
``Yes.'' &
``Leonard's work is nearly as amazing as third graders growing lima beans in wet paper towels.'' \\
\midrule
1\_5679 & Neg & Pos &
``The organism responsible for Michelob Lite.''\newline
``Is there something wrong with your neck?''\newline
``It's a little stiff.''\newline
``What a remarkably fragile structure to support such a valuable payload.''\newline
``Not unlike balancing a Faberg\'e egg on a pixie stick.''\newline
``Have you considered massage?'' &
``I'd like to respond to that sarcastically.''\newline
``Yes, I relish the thought of a stranger covering my body with oil and rubbing it.'' \\
\bottomrule
\end{tabularx}
\end{table}

\section{Case Study}

We present two representative cases to further illustrate the advantages of CUCI-Net in context-aware cue construction and cue-guided multimodal reasoning. Specifically, both cases are correctly classified by CUCI-Net and are selected to exemplify two typical forms of context-dependent multimodal understanding.

\par\noindent
\textbf{Case 1 (MUStARD, hid: 1\_3293): Context-conditioned sarcastic reversal.}
The preceding context starts with a discussion about scientific interests and then shifts to Leonard's physics research, including statements such as ``Actually, my tests of the Bohm quantum interference effect have reached an interesting point,'' ``Right now we're testing the phase shift due to an electric potential,'' and the response ``That's amazing.''
The target utterance then concludes with ``Leonard's work is nearly as amazing as third graders growing lima beans in wet paper towels.''
This sample is difficult to interpret from the target utterance alone, because the intended meaning is not triggered by a single lexical marker, but by the sharp mismatch between the seemingly appreciative scientific context and the trivial comparison introduced at the end.
CUCI-Net preserves the context--utterance structure in the first stage, extracts a discrepancy-aware interpretation cue from the contrast between the contextual buildup and the final expression, and then uses this cue to guide the final multimodal interaction.
As a result, the model is better able to recognize that the target utterance is not a literal compliment, but a context-conditioned sarcastic response.

\par\noindent
\textbf{Case 2 (MUStARD, hid: 1\_5679): Pragmatic sarcasm under explicit conversational setup.}
In this example, the context begins with a humorous remark, then moves to a discussion of the speaker's stiff neck, followed by the question ``Have you considered massage?''
The target utterance responds: ``I'd like to respond to that sarcastically. Yes, I relish the thought of a stranger covering my body with oil and rubbing it.''
Although the utterance contains a superficially positive expression such as ``I relish the thought,'' its intended meaning is clearly negative in the given conversational setting.
The effect arises from pragmatic inversion: the literal wording appears favorable, while the surrounding context reveals discomfort and resistance.
For this type of sample, CUCI-Net benefits from explicitly modeling the dependency between the preceding conversational trigger and the current utterance.
The resulting interpretation cue helps the model precisely focus on the mismatch between literal sentiment and intended attitude, thereby greatly supporting more reliable sarcasm recognition.

Overall, these cases suggest that the advantage of CUCI-Net lies not only in learning stronger multimodal representations, but also in explicitly converting context--utterance dependency into an interpretation cue for downstream reasoning.
This is particularly useful when the final meaning is expressed through semantic reversal, pragmatic mismatch, or context-sensitive attitude shift rather than through obvious lexical indicators alone.

\section{Acknowledgement}
This work was supported by the Science and Technology Innovation 2025 Major Project of Ningbo City (Grant No. 2023Z130)

\bibliographystyle{ACM-Reference-Format}
\bibliography{CUCIN}

\end{document}